\documentclass[9pt,conference]{IEEEtran}

\usepackage[preprint]{waspaa25}

\usepackage{bm} 
\usepackage{subcaption, multirow}
\usepackage{gensymb}

\newcommand{\CH}{\mathrm{ch}}
\newcommand{\Left}{\mathrm{left}}
\newcommand{\Right}{\mathrm{right}}

\title{Head-Related Transfer Function Individualization Using\\Anthropometric Features and Spatially Independent Latent Representation}

\name{Ryan Niu$^{1,2}$,
      Shoichi Koyama$^{1}$,
      Tomohiko Nakamura$^{3}$
      \thanks{This work was supported by the JSPS KAKENHI, Grant Number 23K24864, JST FOREST Program, Grant Number JPMJFR216M.}}
\address{$^{1}$National Institute of Informatics, Tokyo, Japan \;
$^{2}$University of Southern California, Los Angeles, CA, USA\\
$^{3}$National Institute of Advanced Industrial Science and Technology, Tokyo, Japan
}

\begin{document}

\maketitle

\begin{abstract}
A method for head-related transfer function (HRTF) individualization from the subject's anthropometric parameters is proposed. Due to the high cost of measurement, the number of subjects included in many HRTF datasets is limited, and the number of those that include anthropometric parameters is even smaller. Therefore, HRTF individualization based on deep neural networks (DNNs) is a challenging task. We propose a HRTF individualization method using the latent representation of HRTF magnitude obtained through an autoencoder conditioned on sound source positions, which makes it possible to combine multiple HRTF datasets with different measured source positions, and makes the network training tractable by reducing the number of parameters to be estimated from anthropometric parameters. Experimental evaluation shows that high estimation accuracy is achieved by the proposed method, compared to current DNN-based methods. 
\end{abstract}

\section{Introduction}
\label{sec:intro}

One of the most promising techniques for high-fidelity spatial audio reproduction for virtual/augmented reality applications is binaural rendering, which aims to reproduce sound in both ears using headphones. Binaural signals are synthesized by using head-related transfer functions (HRTFs), which represent the acoustic transfer characteristics from a sound source to both ears. HRTFs depend on the anatomical characteristics of the individual, such as the shape of the pinna, resulting in a significant variety of HRTFs across individuals~\cite{Blauert97}.
Because individual HRTFs uniquely shape sound perception, their use is necessary for reproducing high-fidelity binaural signals~\cite{Wenzel+93,Moller+96}.
However, measuring the HRTFs through the sequential recording of impulse responses is impractical due to the significant economic and time costs involved.



One strategy for obtaining an individual's HRTF in a simple and efficient manner is to individualize the HRTF using anthropometric parameters measured for head, torso, and pinnae~\cite{Zotkin+03,Fantini:IEEE_OJSP2025}. These parameters can be either measured manually or extracted from images or 3D scans, making them easier to obtained compared to each individual's HRTFs themselves. Therefore, there have been many studies on individualizing HRTFs from anthropometric parameters. Earlier studies include methods based on numerical simulation and methods for selecting a similar HRTF~\cite{Guezenoc+20}. However, numerical-simulation-based methods require high-resolution 3D scans, and selection-based methods require a sufficiently representative database for the selection. 
In recent years, machine learning (ML)-based techniques, in particular deep neural network (DNN)-based techniques, have been investigated to overcome the drawbacks of the conventional methods~\cite{Martinez:Acoustics2023,Wang:AESconv2021,Xi:WASPAA2021,Hu+08,Meng:ICME2018,Zhang:IEEE_ACM_J_ASLP2020,Grijalva:ICASSP2014,Chen:ICASSP2019,Lu:IEEE_ACCESS2021,Yao:JASA_EL2022,Miccini:IEEE_VRW2021}. DNN-based techniques are expected to demonstrate high performance in HRTF individualization by learning a highly nonlinear relationship between anthropometric parameters and HRTF, owing to their high representational power. 

One of the challenges in applying DNN-based techniques to HRTF individualization is the small amount of data. Due to the high cost of HRTF measurement, the number of subjects in many measured HRTF datasets is often limited to about 100--200, and the number of those with anthropometric data is even smaller. Therefore, it is desirable to combine multiple datasets, yet this is not straightforward because the measured source positions often differ depending on the dataset. Since many DNN-based methods require that the source positions in the training data are all the same, compensation for the source position mismatches is essential to combine multiple datasets.
In addition, anthropometric parameters are typically much lower in dimension than HRTFs.
It is necessary to estimate high-dimensional data from low-dimensional data, which poses a high-dimensional regression problem from limited data resources. 

We propose an HRTF individualization method that estimates the log-magnitude of HRTFs from anthropometric parameters through their latent representations independent of source positions. The latent representations of HRTFs are obtained using an autoencoder that was first proposed for HRTF upsampling by Ito~\textit{et al.}~\cite{Ito:IWAENC2022} and extended in \cite{Ito:IEEE_OJSP2025}. 
The weights of this autoencoder are conditioned on source positions and frequency, while the latent space is independent of source position. This enables us to handle HRTFs measured at different source positions.
In the proposed method, the anthropometric parameters are fed into a DNN trained to estimate the corresponding latent representations of the autoencoder. These latent representations are then passed to the decoder to obtain HRTFs.

The advantages of our proposed method are twofold.
First, it is possible to combine multiple datasets owing to the source position independence of the latent space.
Second, network training becomes more tractable because the HRTF individualization problem is reduced to the problem of estimating subject-specific latent representations, while the autoencoder accounts for the HRTF characteristics across source positions. This enables effective prediction from low-dimensional anthropometric parameters. 
Two neural network architectures are investigated to estimate the latent representations from anthropometric parameters. Their effectiveness is evaluated through experimental comparison with DNN-based HRTF individualization methods that directly estimate HRTFs from anthropometric parameters.
The code for reproducing the results is available at \url{https://github.com/skoyamalab/HRTFLatentIndividualization}.

\section{Problem Statement}
\label{sec:prob}

Let $h_{s,b,\CH, l}\in\mathbb{C}$ denote the HRTF for subject $s\in\{1, \ldots, S\}$, source position $b\in\{1,\ldots,B\}$, channel $\CH \in \{\Left, \Right\}$, and frequency bin $l \in \{1, \ldots, L\}$, where $S$ is the number of subjects, $B$ is the number of source positions, and $L$ is the number of frequency bins. The $b$th source position is defined as $\bm{x}_b\in\mathbb{R}^3$. Each subject's anthropometric parameters are denoted by $\bm{\alpha}_{s,\CH}\in\mathbb{R}^J$, where $J$ is the number of parameters per channel. The objective of the HRTF individualization problem is to estimate $h_{s,b,\CH, l}$ from $\bm{\alpha}_{s,\CH}$ for the target source positions. Since the phase of $h_{s,b,\CH, l}$ is usually obtained by minimum phase reconstruction from the magnitude of $h_{s,b,\CH, l}$, we focus on estimating the log-magnitude of $h_{s,b,\CH, l}$, i.e., $\mathring{h}_{s,b,\CH, l} := 20 \log_{10} |h_{s,b,\CH, l}|$, from $\bm{\alpha}_{s,\CH}$. The head-related impulse response (HRIR) is obtained from $h_{s,b,\CH, l}$ with the ITD estimated separately. 

The anthropometric parameters are included in several HRTF datasets, e.g., CIPIC~\cite{Algazi+01}, HUTUBS~\cite{Brinkmann+19b}, and CHEDAR~\cite{Ghorbal+20}.
We adopted $J=23$ shared parameters between both CIPIC and HUTUBS, which are head width, height, and depth ($x_1$, $x_2$, and $x_3$), pinna offset down and back ($x_4$ and $x_5$), neck width, height, and depth ($x_6$, $x_7$, and $x_8$), torso top width ($x_9$), shoulder width ($x_{12}$), height ($x_{14}$), head circumference ($x_{16}$), shoulder circumference ($x_{17}$), cavum concha height ($d_1$), cymba concha height and width ($d_2$ and $d_3$), fossa height ($d_4$), pinna height and width ($d_5$ and $d_6$), intertragal incisure width ($d_7$), cavum concha depth ($d_8$), pinna rotation angle ($\theta_1$), and pinna flare angle ($\theta_2$), for each ear's $\bm{\alpha}_{s,\CH}$. 

\section{Prior Work}
\label{sec:prior}

Many studies have been conducted on HRTF individualization using anthropometric parameters~\cite{Fantini:IEEE_OJSP2025}. Linear regression and sparse regression have been used in the past~\cite{Nishino:AppAcoust2007,Grindlay:ICASSP2007,Bilinski:ICASSP2014,Zhu:IC3D2017,Iida:AppAcoust2019,Wang:AppSci2019}. In some cases, HRTF magnitude is estimated directly from anthropometric parameters~\cite{Iida:AppAcoust2019}, but in most cases, dimensionality reduction of HRTF magnitude is performed using spherical harmonic transform (SHT), principal component analysis (PCA), or other methods~\cite{Nishino:AppAcoust2007,Grindlay:ICASSP2007,Wang:AppSci2019}.

In recent years, DNNs have been widely used due to their high performance in various tasks. Even in DNN-based methods, dimensionality reduction is often performed as a preprocessing, with techniques such as SHT~\cite{Wang:AESconv2021,Xi:WASPAA2021}, PCA~\cite{Hu+08,Meng:ICME2018,Zhang:IEEE_ACM_J_ASLP2020}, Isomap~\cite{Grijalva:ICASSP2014}, and (variational) autoencoders~\cite{Chen:ICASSP2019,Lu:IEEE_ACCESS2021,Yao:JASA_EL2022,Miccini:IEEE_VRW2021} being used. Alternatively, methods that directly estimate HRTF or HRIR also exist~\cite{Zhang:CAAI_IntellTech2023,Martinez:Acoustics2023,Sanchez:ICASSP2025}. 

Obtaining a latent representation of HRTFs through nonlinear transformations such as (variational) autoencoders is expected to be effective in HRTF individualization because it more effectively models the complex nonlinear relationships between anthropometric parameters and HRTFs and extracts representative features while reducing overfitting due to limited data, compared to linear transformations such as SHT and PCA. However, current methods encode HRTFs into latent representations depending on source positions, making it difficult to combine datasets with different measurement positions. 

\section{Proposed Method}
\label{sec:prop}

\begin{figure}[t]
    \centering
    \begin{subfigure}{\columnwidth}
        \centering
        \includegraphics[width=0.83\columnwidth]{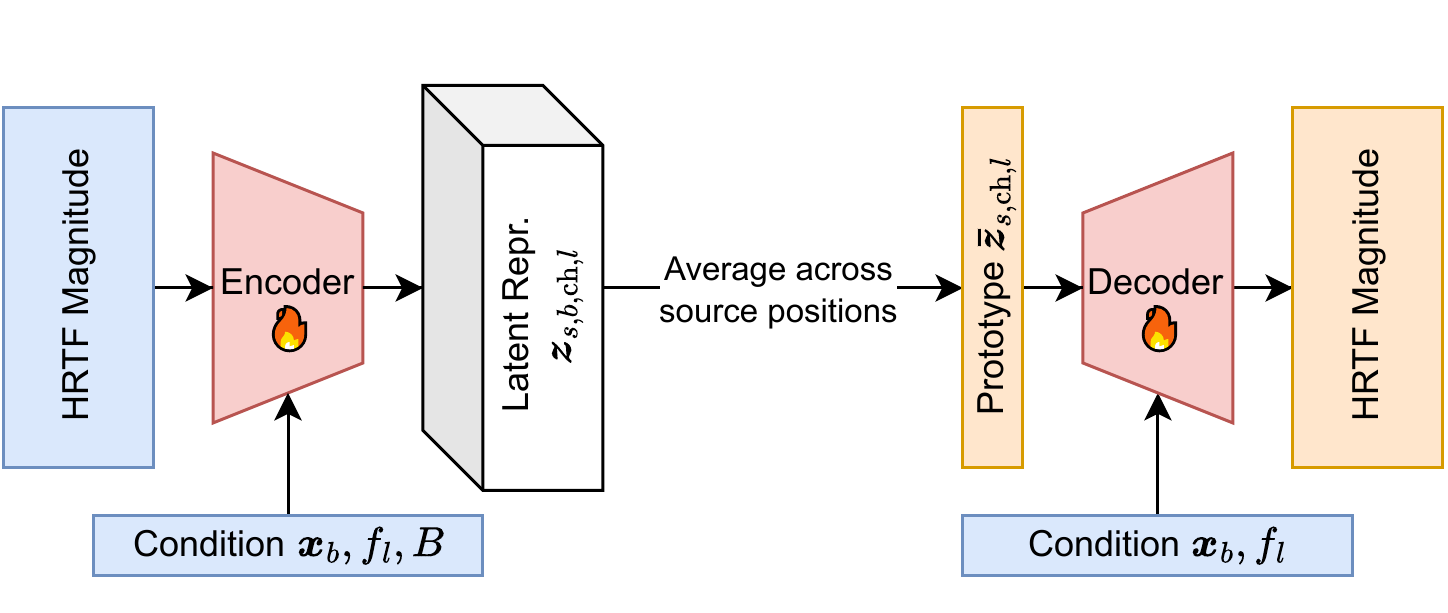}
        \vspace{-3mm}
        \caption{Autoencoder pretraining.}
    \end{subfigure}
    \\
    \vspace{3mm}
    \begin{subfigure}{\columnwidth}
        \centering
        \includegraphics[width=0.65\columnwidth]{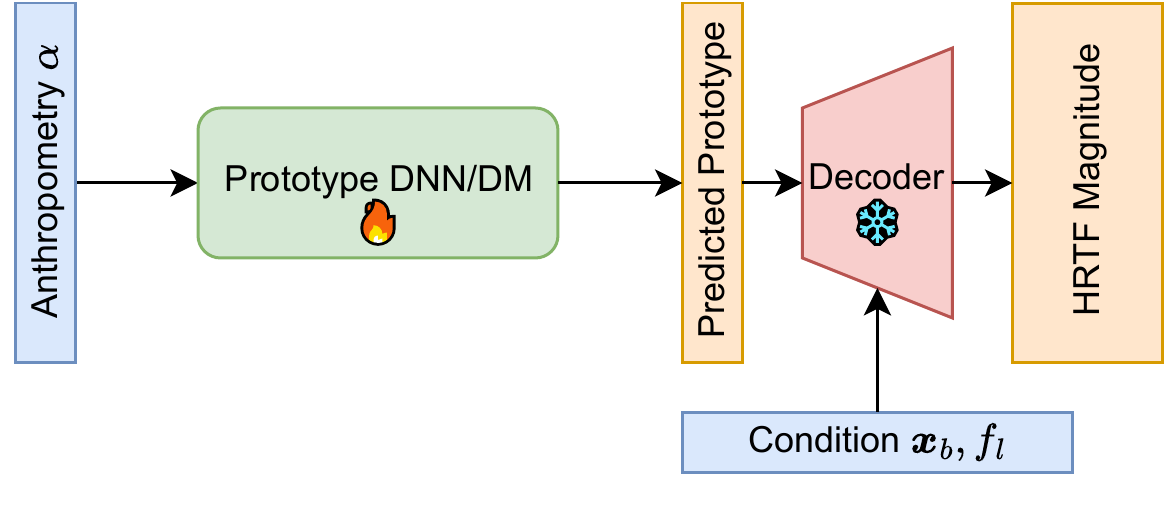} 
        \vspace{-3mm}
        \caption{Prototype estimation from anthropometric parameters.}
    \end{subfigure}
    \caption{Proposed approach for HRTF individualization from anthropometric parameters using source-position-independent latent space.}
    \label{fig:concept}
\end{figure}

We propose a HRTF individualization method based on a latent representation independent of source positions obtained by an autoencoder conditioned on source positions and frequency. The schematic diagram of our proposed method is shown in Fig.~\ref{fig:concept}. 
The proposed network generates individualized HRTFs in two steps. First, a source-position-independent latent representation, referred to as a \emph{prototype}, is estimated from anthropometric parameters. Second, this prototype is converted into the log-magnitude of the HRTF using a pretrained autoencoder.
This two-stage approach explicitly decomposes individual variation and directional dependence in HRTFs, enabling more efficient and generalizable estimation.
In the subsequent sections, we first detail the autoencoder that performs HRTF reconstruction in the second step, and then present the neural network that maps anthropometric parameters to prototypes in the first step.

\subsection{Autoencoder conditioned on source positions and frequencies}

The autoencoder takes the log-magnitude of the HRTF of size $B \times 2L$ for each subject $s$ as input, where the left and right channels are concatenated along the frequency dimension. It encodes the HRTF magnitude into latent representations $\bm{z}_{s,b,\CH,l}$ of size $B \times 2L \times D$, where $D$ is the number of latent representations. The encoder is conditioned on the source position and frequency. Then, the latent representations are averaged along the source position dimension to obtain source-position-independent prototypes $\bar{\bm{z}}_{s,\CH,l}$ of size $2L \times D$. The decoder reconstructs the log-magnitude of HRTF of size $B \times 2L$ from the prototypes, conditioned on the source position and frequency. Importantly, since the prototypes are source-position-independent, the source positions used in the decoder can differ from those used in the encoder. Owing to this characteristic, this autoencoder can handle HRTF data measured at different source positions in a single network.

The specific network architecture of the autoencoder is the same as that described in \cite{Ito:IWAENC2022,Ito:IEEE_OJSP2025}. The encoder and decoder consist of layer normalization, Mish nonlinearity~\cite{Misra19}, and two hyperlinear layers~\cite{Ha+17}. The weights and biases of the hyperlinear layers are obtained from the conditioning vector of source position and frequency through the weight/bias generator. The weight/bias generator of the encoder and decoder consists of Fourier feature mapping (FFM)~\cite{Tancik+20}, layer normalization, Mish nonlinearity, and fully-connected linear layers. The details of FFM will be shown later in \cref{sec:prototype}. The difference from \cite{Ito:IWAENC2022} is the use of frequency conditioning and FFM as in \cite{Ito:IEEE_OJSP2025}, but ITD is not predicted in this paper. 

After pretraining, the frozen decoder serves as a generator that maps prototypes to HRTF magnitudes in the proposed network.

\subsection{Prototype estimation from anthropometric parameters} \label{sec:prototype}
To estimate the source-position-independent prototypes $\bar{\bm{z}}_{s,\CH,l}$ from the anthropometric parameters $\bm{\alpha}_{s,\CH}$, we propose two neural network architectures: \textit{prototype DNN} and \textit{prototype diffusion model (DM)}.

\begin{figure}
\centering
\hspace{-5mm} 
\subfloat[Prototype DNN.]{
    \vspace{-4mm}
    \includegraphics[width=0.51\columnwidth]{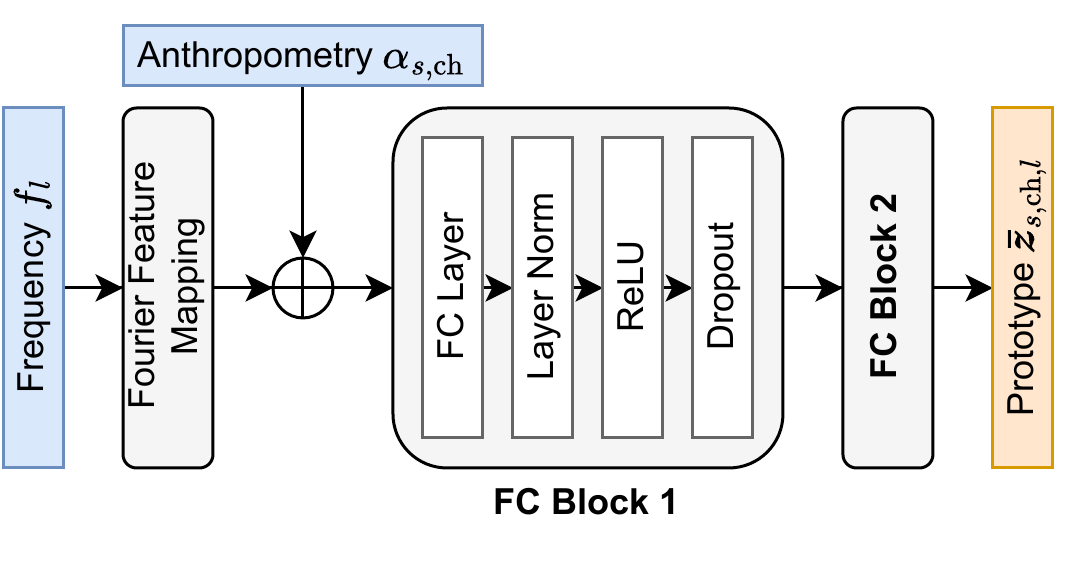}
    \label{fig:ldnn}
}
\\
\vspace{3mm}
\centering
\hspace{-5mm} 
\subfloat[Prototype DM.]{
    \vspace{-2mm}
    \includegraphics[width=\columnwidth]{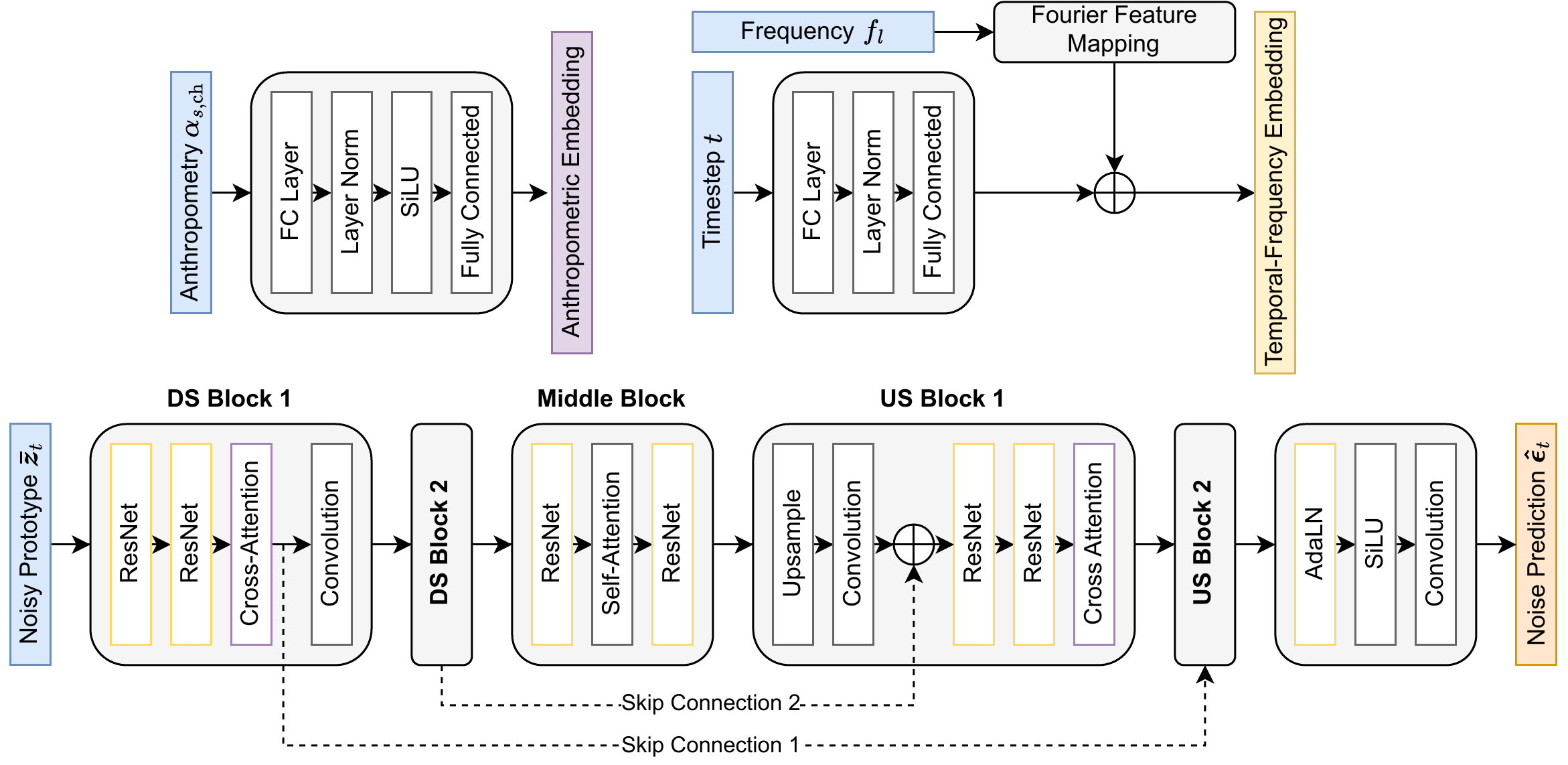}
     \label{fig:ldm}
 }
 \caption{Proposed network architectures for prototype estimation, where $\oplus$ denotes the concatenation of two inputs.}
 \label{fig:proposed_networks}
\end{figure}

\smallskip
\noindent \textbf{Prototype DNN:} 
\cref{fig:ldnn} shows the network architecture of the prototype DNN.
It takes $\bm{\alpha}_{s,\CH}$ and frequency $f_l$ to predict the prototype $\bar{\bm{z}}_{s,\CH,l}$ generated from encoding the target HRTF.
This network consists of two fully-connected (FC) blocks, each consisting of a FC layer, a layer normalization, a rectified linear unit (ReLU) nonlinearity, and a dropout layer.
Before concatenation with the anthropometric features, the frequency $f_l$ is first normalized as $f_l/f_{\text{max}}$, where $f_{\text{max}}$ is the maximum frequency in the data.
The normalized frequency is then passed through the FFM~\cite{Tancik+20}.
As in~\cite{Ito:IWAENC2022}, the FFM maps $f_l$ to a $2K$-dimensional embedding:
$\bm{\phi}(f_l)=[\sin(2\pi \kappa_1 f_l),\cos(2\pi\kappa_1 f_l),\ldots,\sin(2\pi \kappa_K f_l),\cos(2\pi\kappa_{K} f_l)]^\top$
where $\kappa_k$ ($k=1,\ldots,K$) are trainable parameters initialized by sampling from a standard normal distribution.
This network is trained to minimize mean squared errors (MSE) between predicted and ground-truth prototypes.

\smallskip
\noindent \textbf{Prototype DM:}
The second network architecture is based on the latent diffusion model\cite{Rombach:CVPR2022}, which has shown strong performance in image and audio generation tasks \cite{Yang:ACMCS2023}. It combines an autoencoder with a conditional denoising diffusion implicit model (DDIM) \cite{Song:ICLR2021}, a deterministic and accelerated DM variant.

DDIM generates a desired latent representation by progressively denoising Gaussian noise, guided by conditioning inputs and a neural network trained to predict the noise components.
In our implementation, a one-dimensional (1D) U-Net is trained using a MSE loss to estimate the noise component added to the ground-truth prototype.
The training follows the standard DDIM formulation (see \cite{Rombach:CVPR2022,Song:ICLR2021} for mathematical details), where anthropometric features serve as conditioning inputs.
To enhance conditional fidelity during inference, we apply classifier-free guidance (CFG) \cite{ho:2022}, where noise predictions from both conditional and unconditional inputs are interpolated using a scale parameter $w$.
The sampling process also incorporates the DDIM parameter $\eta$, which controls the level of stochasticity during generation: $\eta = 0$ corresponds to a fully deterministic process, while $\eta > 0$ introduces randomness into each denoising step.
The predicted prototype is then decoded to reconstruct the HRTF magnitude.

\cref{fig:ldm} shows our U-Net architecture, which comprises two down-sampling (DS) blocks, a middle block, two up-sampling (US) blocks, and an output block.
For conditional inputs, the network includes an anthropometry embedding block and a timestep-and-frequency embedding block.
The DS and US blocks consist of ResNet blocks with adaptive layer normalization (AdaLN) \cite{xu:2019}, along with cross-attention and self-attention modules. The network includes 1D convolutional layers applied along the frequency axis, both inside and outside the ResNet blocks. Each ResNet block consists of two sequential sub-blocks: the first includes AdaLN, a SiLU activation, and a 1D convolutional layer; the second mirrors this structure but adds a dropout layer for regularization. A residual connection adds the input of the ResNet to its output. The cross-attention blocks are conditioned on anthropometric embeddings, while the AdaLN layers are modulated by the combined timestep and frequency embeddings.


\section{Experimental Evaluation}
\label{sec:exp}


\begin{table}[t]
    \centering
    \caption{Datasets used in HRTF individualization experiments}
    \label{tab:datasets}
    \begin{tabular}{c|cc|cc} \toprule
    \multirow{2}{*}{Dataset} & \# Source & Source & \multicolumn{2}{c}{\# Subjects} \\
    & pos. $B$ & dist. $r$ (m) & Train & Test \\ \midrule
    CIPIC & 1250 & 1.00 & 30 & 5 \\
    HUTUBS & 440 & 1.47 & 85 & 6 \\ 
       \bottomrule
    \end{tabular}
\end{table}


\begin{figure*}[t!]
    \centering
    \begin{subfigure}{0.3\textwidth}
        \centering
        \centerline{\includegraphics[width=\columnwidth]{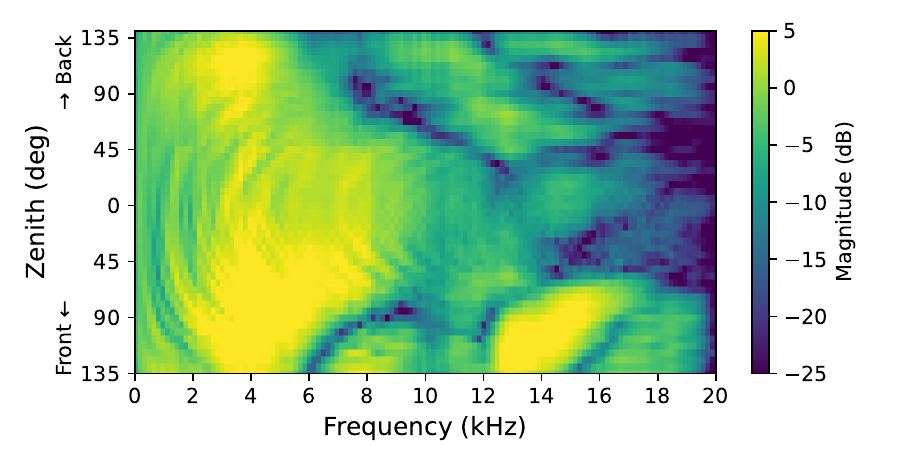}}
        \caption{Ground truth}\label{fig:a}
    \end{subfigure}
    \begin{subfigure}{0.3\textwidth}
        \centering
        \centerline{\includegraphics[width=\columnwidth]{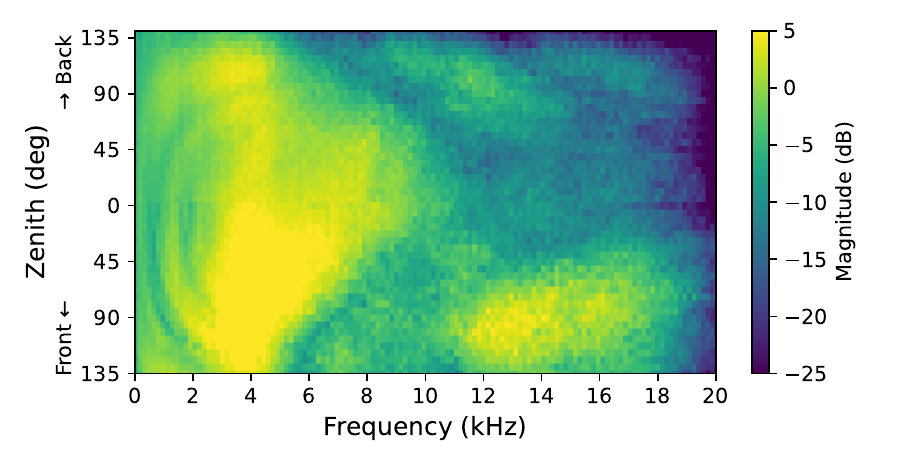}}
        \caption{HRTF DNN}\label{fig:b}
    \end{subfigure}
    \begin{subfigure}{0.3\textwidth}
        \centering
        \centerline{\includegraphics[width=\columnwidth]{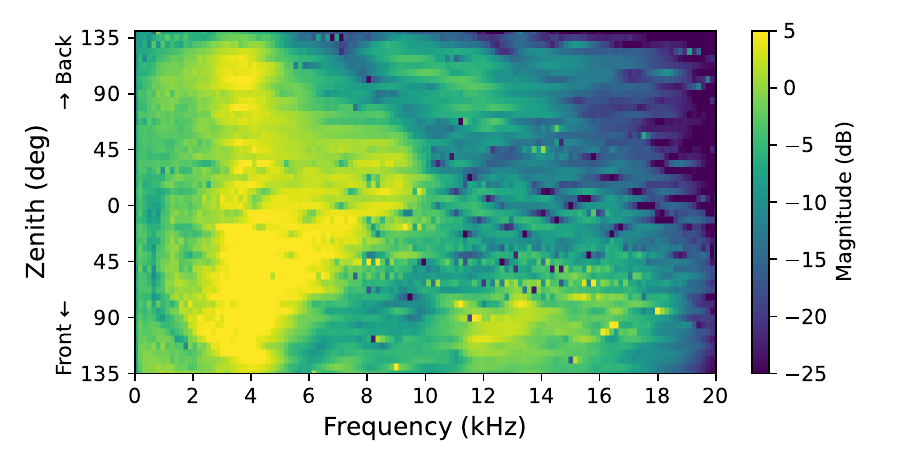}}
        \caption{HRTF DM}\label{fig:c}
    \end{subfigure}
    \begin{subfigure}{0.3\textwidth}
        \centering
        \centerline{\includegraphics[width=\columnwidth]{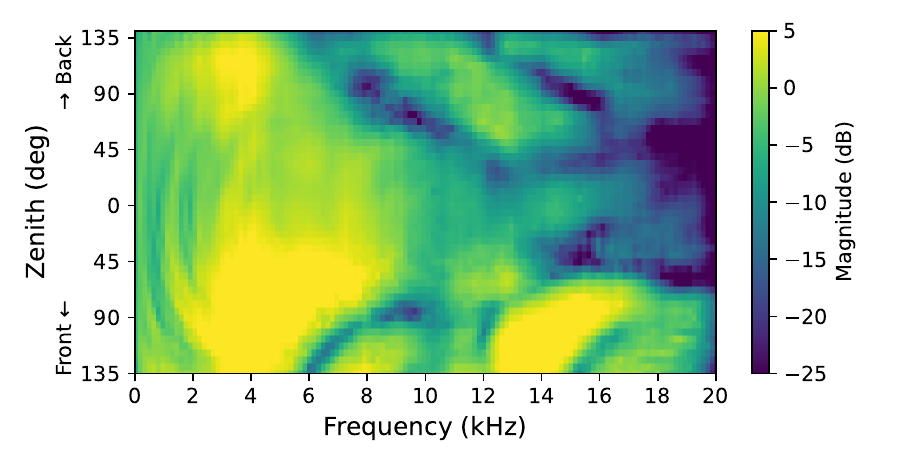}}
        \caption{Autoencoder reconstruction}\label{fig:d}
    \end{subfigure}
    \begin{subfigure}{0.3\textwidth}
        \centering
        \centerline{\includegraphics[width=\columnwidth]{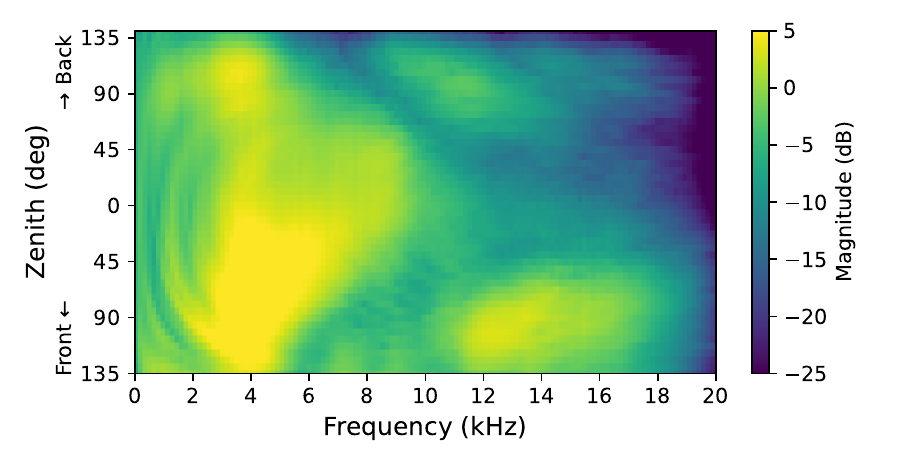}}
        \caption{Proposed prototype DNN}\label{fig:e}
    \end{subfigure}
    \begin{subfigure}{0.3\textwidth}
        \centering
        \centerline{\includegraphics[width=\columnwidth]{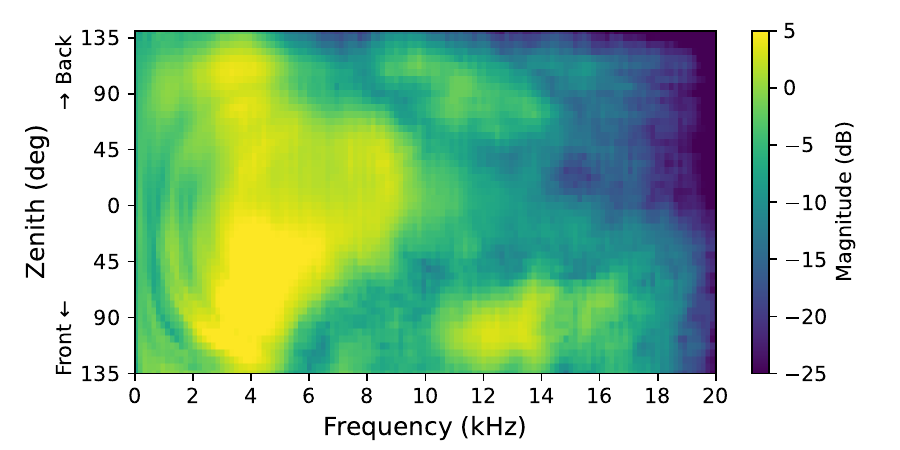}}
        \caption{Proposed prototype DM}\label{fig:f}
    \end{subfigure}
    \caption{Predicted HRTF magnitudes from proposed and baseline models, compared with ground-truth and reconstructed HRTF magnitude from autoencoder. The sample is from the CIPIC dataset. When possible, both datasets (CIPIC + HUTUBS) were used to train the model.} 
    \label{fig:plots}
\end{figure*}

\subsection{Dataset}
To evaluate the effectiveness of the proposed approach, we conducted HRTF individualization experiments using the CIPIC\cite{Algazi+01} and HUTUBS\cite{Brinkmann+19b} datasets.
\cref{tab:datasets} shows their characteristics and the train/test splits used in the experiments.
Since the autoencoder does not require anthropometric information, we additionally included 10 and 3 subjects without complete anthropometric parameters for its pretraining on the CIPIC and HUTUBS datasets, respectively. Importantly, both the autoencoder and the individualization networks were trained on the same data subsets: when using only CIPIC, both components were trained on CIPIC; when using the merged dataset, both were trained on the combined CIPIC and HUTUBS data.



Prior to inputting to the neural networks, the HRTF magnitudes, anthropometric features, and prototypes were normalized using z-score normalization (zero mean and unit variance). The normalization statistics were computed solely from the training data. HRTF magnitudes were normalized separately for each dataset, whereas anthropometric features and prototypes were normalized based on the entire training set corresponding to each experimental configuration.

\subsection{Compared methods}
\subsubsection{Proposed networks}
For the prototype DNN and DM, we first pretrained an autoencoder using $L=128$ frequency bins (0 to $f_{\text{max}} = 20$ kHz) and a latent dimension of $D=64$. It was trained to minimize the mean log-spectral distortion (LSD):
\begin{equation}
\mathrm{LSD} = \sum_{s,b,\CH} \sqrt{ \frac{1}{L} \sum_l \left( \hat{\mathring{h}}_{s,b,\CH,l} - \mathring{h}_{s,b,\CH,l} \right)^2 }.
\end{equation}
For FFM, we set $K=8$ for the autoencoder and $K=16$ for the individualization networks.
Both FC layers in the prototype DNN were set to 128 units, and a dropout of 0.5 was used to prevent overfitting.
As for the prototype DM, the 1D convolutional layers inside the network all used a kernel size of 3 and padding of 1. Its stride was set to 2 in the DS blocks for halving sequence length, while the remainder used a stride of 1. The US layers performed linear interpolation with a scale factor of 2 prior to convolution. All attention mechanisms used 8 heads, and a dropout rate of 0.15 was applied within the ResNet and attention layers.
The channel depths of the U-Net were 64 (initial), 128 (after second DS block), and 256 (middle), then symmetrically decreasing back to 64 before the output projection.
Finally, the anthropometric and timestep-and-frequency embedding dimensions were 32 and 192, respectively. 

\subsubsection{Baseline networks}
We compared the two proposed networks with two baseline networks that directly estimate $\mathring{h}_{s,b,\CH, l}$ from $\bm{\alpha}_{s,\CH}$: \textit{HRTF DNN} and \textit{HRTF DM}, which can be regarded as variants of the methods proposed in \cite{Martinez:Acoustics2023} and \cite{Sanchez:ICASSP2025}, respectively.

For a fair comparison, the HRTF DNN mirrors the prototype DNN architecture, but omits frequency conditioning and takes only $\bm{\alpha}_{s,\mathrm{ch}}$ as input. The loss function is defined as the LSD between the predicted and ground-truth HRTF magnitudes. Based on preliminary experiments, the first and second FC layers were set to 64 and 512 units, respectively. The output layer was set to $B \times L$ units to match the size of the HRTF magnitude.


The HRTF DM mirrors the prototype DM, but it operates directly in the HRTF magnitude space, with U-Net conditioning shifted from frequency $f_l$ to source positions $\bm{x}_b$.
As in \cite{Sanchez:ICASSP2025}, HRTF magnitudes are treated as $B$ slices of single-channel data with sequence length $L$, enabling compatibility with varying source positions. Source positions, normalized with the source distance $r$ as $\bm{x}_b / r$, are processed by FFM and concatenated with anthropometric embeddings to form the conditional input for the U-Net. The timestep embedding is used independently as an additive positional encoding in the ResNet blocks.
Due to the removal of frequency conditioning, all AdaLN layers are replaced with standard group normalization. ResNet normalization layers use one input/output channel to match the new data structure.

All other hyperparameters of the HRTF DNN and DM were identical to those of their respective prototype counterparts.

\subsubsection{Training and inference settings}
All networks were trained using the AdamW optimizer \cite{Loshchilov:ICLR2019} for up to 300 epochs, with early stopping based on validation loss. The learning rate and weight decay were selected in the range $1 \times 10^{-4}$ to $1 \times 10^{-3}$ for each network.
Learning rate scheduling differed by model: for both DNNs, the learning rate was reduced when the validation loss plateaued; for the DMs, it followed a cosine annealing schedule over epochs.
Hyperparameters were tuned via 5-fold cross-validation on the training data. The models were then retrained on the full training set using the best hyperparameters and evaluated on a fixed hold-out test set.

%


For DMs, the DDIM sampler was used with a linear noise schedule. Training used 1000 timesteps, while inference used 500. A gradient clipping of 1 was also applied during training. During inference, the predicted prototype was clamped to $[-3, 3]$ at each timestep to stabilize generation.
Empirically, we found that the best performance was achieved with DDIM parameters $w = 4$, $\eta = 0.2$ for the prototype DM, and $w = 2$, $\eta = 0.25$ for the HRTF DM.

\subsection{Results}
\begin{table}[t!]
    \centering
    \caption{
        Mean and standard deviation of LSD (dB) across different datasets. Results are reported for both baseline and proposed methods. ${}^+$ indicates models trained jointly on the CIPIC and HUTUBS datasets.
    }
    \label{tab:results}
    \sisetup{
       reset-text-series = false, 
       text-series-to-math = true, 
       mode=text,
       tight-spacing=true,
       table-number-alignment=center
    }
    {
    \tabcolsep=0.66em
    \begin{tabular}{c|cc|cc}
       \toprule
       \multirow{2}{*}{Dataset} & \multicolumn{2}{c|}{Baseline} & \multicolumn{2}{c}{Proposed} \\ 
       & {HRTF DNN}  & {HRTF DM} & {Proto. DNN} & {Proto. DM} \\
       \midrule
       CIPIC & 4.99 $\pm$ 0.23 & 5.67 $\pm$ 0.25 & \textbf{4.97 $\pm$ 0.25} & 5.25 $\pm$ 0.33 \\
       CIPIC${}^+$& N/A & 5.60 $\pm$ 0.19 & \textbf{4.90 $\pm$ 0.29} &  5.07 $\pm$ 0.25 \\
       \midrule
       HUTUBS & 5.09 $\pm$ 0.41 & 5.50 $\pm$ 0.34 & \textbf{5.01 $\pm$ 0.35} & 5.35 $\pm$ 0.36 \\
       HUTUBS${}^+$& N/A & 5.49 $\pm$ 0.35 & \textbf{5.07 $\pm$ 0.34} & 5.21 $\pm$ 0.40 \\
       \bottomrule
    \end{tabular}
    }
\end{table}

\cref{tab:results} shows the LSDs between predicted and ground-truth HRTF magnitudes for each test set.
The proposed networks consistently outperformed the baseline counterparts on both the CIPIC and HUTUBS datasets.
The prototype DNN achieved the best performance in all conditions, while using only 32k parameters compared to 82M for the HRTF DNN on CIPIC.
Even when including the decoder part of the autoencoder, the total number of parameters remains substantially smaller (approximately 260k).
These results demonstrate both the effectiveness and efficiency of the proposed approach.

The HRTF DNN could not be jointly trained on CIPIC and HUTUBS due to incompatible source positions, which prevent direct prediction over a unified direction set.
All other networks benefited from joint training, showing improved or comparable performance.
One exception was the prototype DNN on the HUTUBS test set, where performance slightly decreased; however, it still outperformed all other networks.
Among diffusion models, the prototype DM showed greater gains than the HRTF DM, indicating that using source-position-independent prototypes supports better generalization across datasets.

\cref{fig:plots} shows examples of predicted HRTF magnitudes for a CIPIC test subject, where all networks except the HRTF DNN were trained with both datasets.
As a reference, the figure includes the reconstructed result by the autoencoder.
The proposed networks more closely match the original and reconstructed HRTFs, whereas the baseline predictions contain more artifacts and noise.
These observations support the qualitative effectiveness of the proposed approach.




\section{Conclusion}
\label{sec:concl}
We proposed an individualization method that estimates HRTF magnitudes through a source-position-independent latent space.
The method first trains an autoencoder conditioned on source position and frequency to learn latent representations of HRTFs and derive source-position-independent, subject-specific prototypes.
A separate network is then trained to estimate the prototypes from anthropometric parameters.
We investigated both DNNs and DMs operating in the latent space.
Experiments showed that the proposed networks outperformed baseline models that directly estimate HRTF magnitudes, demonstrating the effectiveness of our approach.

\clearpage
\bibliographystyle{IEEEtran}
\bibliography{str_def_abrv,skoyamalab_en,refs25}








\end{document}